\documentclass[aps,pra,amsmath,amssymb,nofootinbib,preprint]{revtex4}
\usepackage{graphicx}

\def\LK{\left(}
\def\RK{\right)}

\def\be{\begin{equation}}
\def\ee{\end{equation}}
\def\bea{\begin{eqnarray}}
\def\eea{\end{eqnarray}}

\def\emph#1{{\bf{#1}}}

\begin{document}
\title{Remark on internal relaxation in large number qubit systems}
\author{Martin Jan{\ss}en}
\email{mj@thp.uni-koeln.de}
\affiliation{Institut~f\"ur~Theoretische~Physik, Universit\"at~zu~K\"oln\\ Z\"ulpicher~Str.~77, D-50937~K\"oln}
\date{2026-11-10}
\begin{abstract}
For a time  $t_{c}$  of  calculation in an $N$ qubit quantum computer that is much shorter than the intrinsic Heisenberg time $t_H$, $t_c\ll t_H$, the dynamics of the system is effectively irreversible and the couplings between the qubits give rise to a relaxation time scale $\tau_r$. If  $t_H\gg t_c >\sim \tau_r$ one looses quantum coherence exponentially fast. Since $t_H$ scales exponentially with $N$ there is a threshold $N_c$ such that for $N\gg N_c$ one must be aware of possible intrinsic relaxation during calculation. As an order of magnitude estimation we find $N_c\approx \ln(\tau_r/T)$ where  $T$ is a typical system specific oscillation time of a single qubit.  $\tau_r$ can be estimated as the typical inverse transition rate between two parts of the system, one of which is treated as an environment of the other.
\end{abstract}
\maketitle

A quantum computer (see e.g~\cite{Raseena}) uses a  number $N$ of  coupled qubits for a calculation over a time $t_c$. The actual qubit system may contain more than $N$ qubits, but perhaps  not all of them are involved in the calculation. For simplicity we assume that the total system consists of only those $N$ qubits being an essential part of the calculation. An initial state is prepared at $t=0$ and  at $t=t_c$ a measurement is made. Up to $t_c$ the time evolution of the $N$ qubit has to be unitary to keep the information within the system available. In particular, the dynamics has to be reversible. 
There are a number of obstacles that have to be overcome in quantum computing. We mention error correction and precise implementation of desired dynamical steps for the calculation. Here we only address the influence of relaxation breaking reversibility. As discussed in \cite{HajduJanßen2026} every system behaves essentially irreversible on a time scale which is much shorter than the time scale $t_H$ of the system that is needed  to resolve its discrete energy spectrum. On time scales much shorter than $t_H$ the spectrum cannot be resolved as sharp values and it is best modeled by a continuous spectral density. $t_H$ is often denoted as Heisenberg time of the system. If the system is treated as open (i.e.~coupled to  some environment) $t_H$ is the Heisenberg time of the total closed system  consisting of the open system plus environment. Under this condition the irreversibility becomes evident by the emergence of a relaxator in the system's Liouville operator \cite{HajduJanßen2026} and it results in relaxation with  relaxation  time $\tau_r$, which is specific for the open system.  After time $\tau_r$ states of the system relax to a stationary state situation\footnote{This behavior is generic, exceptions are possible in case of degenerate zero modes of the Liouville \cite{HajduJanßen2026}  due to e.g. symmetries or many body localization.}. Thus, on time scales smaller than the Heisenberg time and larger than the relaxation time the system's state shows relaxation and cannot be described  by a reversible dynamics. Typically, the Heisenberg time grows exponentially with the number of constituents of the corresponding total system. For a large isolated system one can adopt the open system view by considering two parts of it, one of which acts as the environment of the other. Intrinsic relaxation in large  isolated systems is  unavoidable on time scales much smaller than the Heisenberg time\footnote{This is also the area of thermodynamics and entropy considerations.}.

These findings of \cite{HajduJanßen2026} have an effect also for the $N$ qubit system of a quantum computer.
To this end we investigate the typical exponential scaling with $N$ of the Heisenberg time $t_H$ and parametrize it as follows
\be  \label{1}
t_H (N) = T e^{b(N-1)} \, , \, t_H(1)=T\, ,
\ee
where  $T$ is a typical single qubit oscillation time scale, inverse to the single qubit energy separation\footnote{We use units in which $\hbar=2\pi$.}, and $b$ is some system specific real positive number of ${\cal O} (1)$. 
$t_c=t_H(N_c)$ defines a number $N_c$ by
\be\label{2}
N_c = 1+b^{-1}\ln(t_c/T)
\ee 

Since $t_H/ t_c = e^{b(N-N_c)}$, for $N\gg N_c$, the condition $t_H \gg t_c$  is fulfilled with ease. According to \cite{HajduJanßen2026} a relaxation time $\tau_r$ emerges generically in the  dynamics of system states which becomes apparent as exponential relaxation of any initial state to some stationary state with relaxation time factor $e^{-t_c/\tau_r}$. For 
\be
t_H\gg  t_c  >\sim \tau_r \, ,
 \ee
the quantum computer is no longer quantum coherent, as it approaches a stationary state independent of its initial state.

We take $t_c=\tau_r$ as an appropriate threshold criterion such that quantum coherence is no longer guaranteed for
 \be\label{4}
 N\gg N_c \approx \ln (\tau_r/T)\, .
 \ee
In \cite{HajduJanßen2026} 
an order of magnitude estimation for $\tau_r$ was given when it results from a system environment coupling,
\be\label{5}
\tau_r=\tau_{se}\frac{\tau_{se}}{\tau_{e}}\, .
\ee
Here $\tau_e$ is a relaxation time of correlations in the environment and $\tau_{se}$ is the typical inverse rate of transitions between system and environment. 
The threshold number $N_c$ is approximately given as
\be\label{6}
 N_c \approx \ln \LK (\tau_{se}/T)(\tau_{se}/\tau_e)\RK\, .
\ee
For short environmental relaxation time and very weak coupling between system and environment we have $\tau_{se}/\tau_e\gg 1$
and the threshold number $N_c$ for coupling between system and environment can be made large by reducing this coupling (increasing $\tau_{se}$).

The isolated system under the criterion $t_c \ll t_H$ will act as its own environment. Formally, one may take two parts of the system, one of which acts as an environment of the other. The corresponding coupling cannot be made arbitrarily small as the qubits have to be coupled over both parts when using both parts for the calculation within $t_c$. In case of  internal relaxation, the order of magnitude estimate of the relaxation time yields a self-consistency equation, as there is only one typical relaxation time $\tau_r$ of system and environment,
$\tau_r=\tau_{c}\frac{\tau_{c}}{\tau_{r}}$.
Here $\tau_c$ is the typical intrinsic inverse transition rate between the two parts of the system. Consistency requires $\tau_r=\tau_{c}$.
Consequently, intrinsic coupling between two parts causes relaxation in each part. This causes breakdown of quantum coherence for active qubit-numbers $N\gg N_c$ with threshold number $N_c$ approximately given as 
\be\label{9}
N_c\approx \ln(\tau_c/T)\, .
\ee
The intrinsic transition rate cannot be made arbitrarily small without loosing the coupling between the parts of the system. Thus, the only way to reach large numbers $N_c$ is by making the energy separation of a single qubit, $T^{-1}$, much much larger than the typical coupling strength, $\tau_c^{-1}$, between two parts of the system. 

The arguments presented here are reformulated as follows.
 One can explore the full spectral content of a system of $N$ coupled qubits and thus its quantum dynamics within the system's specific relaxation time $\tau_r$ if one reduces $N$ to a number $N_c$ such that the reduced system has a Heisenberg time of order $\tau_r$. Since relaxation comes about also by internal coupling in large $N$ qubit systems, one has to face relaxation to stationary situations when $N\gg N_c\approx \ln(\tau_r/T)$ with $T$ a typical oscillation time in a single qubit system and $\tau_r$ an internal relaxation time which can be estimated as a typical inverse transition rate between two parts of the system, one of which acts as the environment of the other. 
In a calculation with $N\gg N_c$ and for $t_c \ll \tau_r$ one will not take notice of relaxation. However, one cannot explore the full system's dynamic, but only a part of up to the order of $N_c$ qubits.

\end{document}